# What does Maxwell's demon want from life?
## When information becomes functional and physical

J. H. van Hateren

Johann Bernouilli Institute for Mathematics and Computer Science, University of Groningen, Groningen, The Netherlands

email: j.h.van.hateren@rug.nl

**Abstract**  Szilard's one-molecule engine is operated by a Maxwell's demon attempting to convert heat to work. It is argued here that when using the demon to relate information to work (Landauer's principle), the demon's goal must be regarded as an implicit assumption that should not be taken for granted. Other demons than the standard one can be constructed depending on their assumed goal. When the demon is purely mechanical, the goal is instilled in it by its design. When the demon is regarded as "intelligent" (Maxwell) and "endowed with free will" (Thomson), it is not so clear whence he could acquire a specific goal. A solution is to subject the demon to evolution by natural selection, which in an extended form can provide the demon with a genuine goal for which work is necessary. This implies that the information the demon acquires and uses for operating the engine becomes functional for the demon, and thereby acquires a real physical status. Such information, called functional information, applies to all life and technology. It should be distinguished from the formal, causally ineffective information that can be defined, more arbitrarily, for any system. Information in general is not physical, only functional information is. A similar functional/formal distinction applies to close associates of information, in particular entropy, probability, and the second law of thermodynamics.

*Keywords*  Landauer's principle; Information; Meaning; Entropy; Probability; Second law of thermodynamics

## 1. Introduction

Maxwell (1871) emphasized the statistical nature of the second law of thermodynamics by proposing a hypothetical little creature, later known as Maxwell's demon, capitalizing on the varying speeds of molecules in a gas. The demon observes the speed of individual gas molecules and operates a little door in the wall separating two volumes of gas. The volumes are initially in thermal equilibrium with each other and with a surrounding constant-temperature reservoir (heat bath). By opening and closing the door at the right times, the demon lets fast molecules go to one of the volumes, and slow molecules to the other. The resulting temperature difference between the volumes could then be used to produce work, with the energy balance maintained by an equivalent amount of heat flowing from the heat bath to the gas. The demon thus appears to convert thermal energy of a system in equilibrium into work, whereas the second law states that such a conversion is not possible, at least not continuously (Jarzynski, 2011).

    Because the second law is well established empirically, with no known exceptions, Maxwell's thought experiment poses a paradox. The history of the attempts to resolve it is long and fascinating (for key papers see Leff & Rex, 2003). Szilard (1929) made important progress by simplifying the problem to a one-molecule gas, thus making it more readily analysable. The demon could now produce work by observing the position of the molecule, as explained in more detail in Section 2. Szilard concluded that the act of measuring the position must require work by the demon, such that the second law holds and the paradox is resolved. Landauer (1961) and Bennett (1982) realized that the key process was not the observation itself, but the fact that a memory needs to be reset if the process is run in a cyclic manner. This resetting (information erasure in Landauer's terms) costs at least as much work as the demon can produce (Landauer's principle). This connection of work with information is now the accepted view of how the paradox is resolved (Maruyama et al., 2009).

Maxwell's demon thus obtained a new function: not as a potential challenge to the second law, but as a means to give a physical interpretation to information. Information is indeed increasingly regarded as an intricate part of physics, the so-called informational turn in physics (Belfer, 2012, 2014; Robinson & Bawden 2014). However, I will argue here that "information is physical" is not true in general, but only for special systems.

This article is not about the validity of the second law, which will be taken for granted for all systems discussed here. Also Landauer's principle relating information to work will not be challenged. But I will argue that connecting information to work through demons like those of Maxwell and Szilard is not unequivocal. The demons are based on an implicit, non-necessary assumption with respect to their goals that determines if and when information can be given a physical interpretation. The article is structured as follows. In Section 2 I first explain Szilard's one-molecule gas and how it is used by the demon. I then argue in Section 3 that the demon has an evil twin brother[1], the anti-demon, who has opposite views on what he should do with the gas. Assuming, with tongue-in-cheek, that the demons are not only lively and intelligent (Maxwell, 1871) but also have a will of their own (Thomson, 1874), Section 4 argues that true goals, such as what to do with a gas, could only arise in evolution by natural selection. This implies that information can only be functional for living systems, with an extension to information in technology, which is functional for humans. Only functional information can be given a physical interpretation in the sense that it really exists (independent of human theorizing). The use of information for analysing the world outside the sphere of life and technology should be regarded as a formal tool only (Section 5). Because information is intimately tied to entropy, probability, and the second law, the universality – but not the objectivity – of these concepts is questioned in Section 6. Section 7 concludes.

## 2. Szilard's incarnation of Maxwell's demon

The one-molecule engine proposed by Szilard (1929) consists of a cylinder that is in thermal equilibrium with a surrounding heat bath and that contains a single molecule of which the position is initially unknown (Fig. 1, left). A demon (not shown) then measures the position of the molecule with sufficient resolution to decide if the molecule is in the left half of the cylinder (upper branch of Fig. 1) or in the right half (lower branch). Subsequently, the demon inserts a piston into the cylinder with a cord, pulley, and weight attached to the left or to the right, depending on which half the molecule is in (middle diagrams). The molecule exerts pressure on the piston (as a result of continual collisions), moving the piston and lifting the weight, thereby producing work. Finally, the demon removes the piston. The molecule is then again moving in the complete volume of the cylinder, with its position unknown (right diagram). It is assumed that all movements, such as of the piston and weight, are frictionless. The state of the system is assumed to change in a reversible way (implying quasi-static changes) by continually adjusting the weight to almost match the (volume-dependent) pressure exerted by the molecule on the piston.

While the molecule is pushing the piston towards one end of the cylinder, it will transfer momentum and loose a little of its kinetic energy with each collision with the piston. But the molecule is not cooling down, because it remains in thermal equilibrium with the cylinder and the heat bath, and thus regains the lost energy, on average. Another way to say this is that during the isothermal expansion of the gas, heat is transferred from the heat bath to the gas. By operating the machine in a cyclic manner, the demon can convert thermal energy from the heat bath into the work required for lifting the weight. The energy stored in the lifted weight could then be utilized for other purposes.

However, the machine cannot work for free. Landauer (1961) argued that cyclic operation requires the demon to reset (erase) a one-bit memory at each cycle. This requires minimally the same amount of work as the gas can deliver ($kT\ln 2$, with $T$ temperature, and $k$ the Boltzmann constant[2]). The state of the memory determines the position, left or right, of the weight attached to the piston. It is assumed that attaching and detaching piston and weight is done without friction and would cost no work. But setting the memory, at the measurement stage of each cycle, would cost work because the previous state of the memory is unknown. That is, unknown to the demon, because he could only know this state by operating a second memory, which would also have to be reset, starting an infinite regress of memories. Conceptually, setting a memory from an unknown state to a specified state is equivalent to first erasing the unknown state (resetting the memory to a default state) and then setting the memory from the default state to the new state. The latter can be done without using work. For example, a memory may consist of a particle in a double potential well with a central potential barrier (Landauer, 1961). Such a memory can be switched from a known state to a newly specified state either by leaving the particle where it is, or by pushing the particle over the barrier. The latter requires work, but this work can be regained when the particle drops back to the bottom of the well on the other side of the barrier.

---

[1] I will refer to lively or living demons as "he", to automatic or non-living demons as "it", and to physicists as "she".
[2] This amount of work follows from the ideal gas law for a single molecule, $PV=kT$, with $P$ pressure and $V$ volume, and integrating $PdV=kTdV/V$ from $V_0/2$ to $V_0$, with $V_0$ the volume of the cylinder.

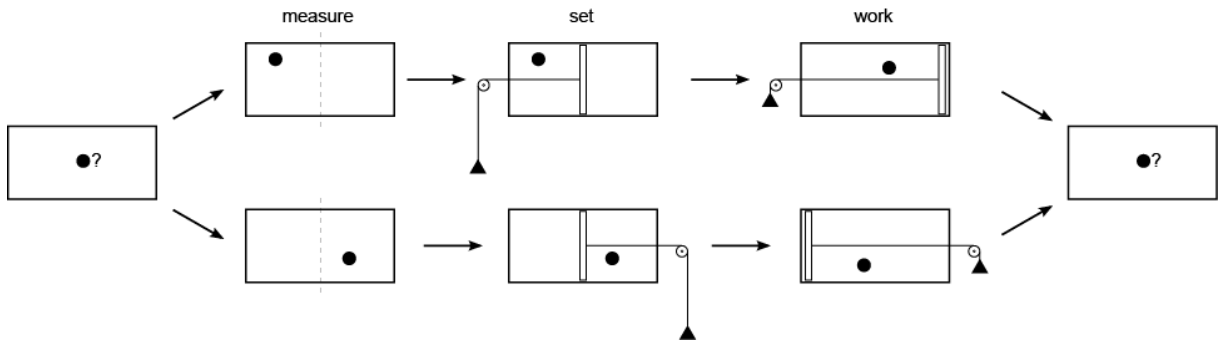

**Fig. 1.** The cyclic behaviour of an engine proposed by Szilard (1929), operated by a demon striving to convert heat into work.

Landauer's principle states that it costs work to erase memory, i.e., to bring a memory from an unknown state into a default one. This can also be formulated in terms of physical (Boltzmann) or informational (Shannon, 1948) entropy. Physical entropy is defined as $k \ln W$, with $W$ the number of states consistent with what is known about the system. Resetting the memory reduces its physical entropy from $k \ln 2$ to $k \ln 1$ (=0), thus $k \ln 2$ needs to be compensated by an equivalent amount of work, $kT \ln 2$, according to the second law (on average). Informational entropy is defined as $-\Sigma_i p_i \ln p_i$, with summation over states $i$ that occur with probability $p_i$. The informational entropy of the memory is reduced from $-\Sigma_i p_i \ln p_i = \ln 2$ (for probabilities $p_1 = 0.5$, $p_2 = 0.5$) to $-\Sigma_i p_i \ln p_i = 0$ (for $p_1 = 1$, $p_2 = 0$). Information is reduction of entropy (with entropy interpreted as uncertainty or missing information), thus by setting the memory $\ln 2$ information is gained[3]. Because $k$ could be replaced by a unitary and dimensionless constant by a suitable redefinition of temperature (Leff, 1999), it can be seen that physical and informational entropy are essentially equivalent here. Hence Landauer's one-liner "information is physical" (Landauer, 1991). On the one hand it costs work to store a bit of information, and on the other hand a stored bit of information can be used to produce work. However, below I will argue that this connection between information and physics is conditional, because it crucially depends on having a specific goal.

## 3. The anti-demon

After each measurement, the demon of Fig. 1 takes the decision to put the weight at the side that will produce work. Apparently, his assumed goal is to produce work. But other demons are possible that perform similar actions, only based on different decisions and different goals. Fig. 2 shows an example of such a demon, called the anti-demon here because his actions seem to be contrary to common sense. After measuring the position of the molecule, this demon decides to put the weight on the "wrong" side. The gas then pushes the piston into the same direction as it is drawn already by the weight. The gas expands now in an irreversible way (not quasi-static). How much energy the molecule can transfer to the piston will depend on details such as the molecular speed and the strength of the gravitational field accelerating the weight. In any case, the molecule can still replenish its energy from the surrounding heat bath, and all available work from the falling weight is eventually converted to heat once the piston collides with the end of the cylinder. Before the next cycle, the anti-demon needs to lift the weight to its original height. The anti-demon is therefore converting work into heat, as quickly as possible given the weight and the pressure of the gas.

---

[3] Information in nat; $\ln 2$ nat corresponds to 1 bit, with bits defined by using the logarithm with base 2 rather than the natural logarithm.

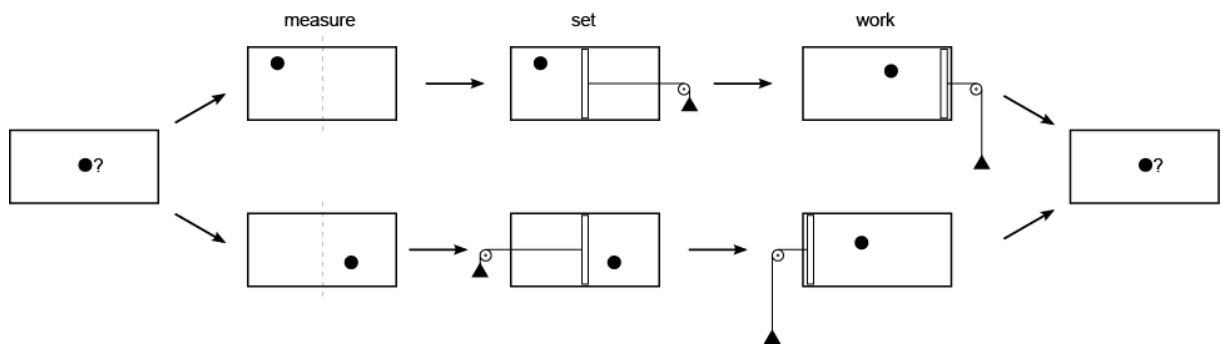

**Fig. 2.** An engine operated by an anti-demon striving to convert work into heat.

The anti-demon converts more work into heat than a mere dropping of the weight would do. Like the demon, the anti-demon still performs a measurement and has to store the position of the molecule in a one-bit memory in order to be able to put the weight on the chosen side. Like the demon, the anti-demon needs to reset this memory at each cycle, and thus also needs work to do that. In other words, the anti-demon is quite effective in destroying work. Moreover, the tight connection between information and work is broken.

Demons intermediate between demon and anti-demon are possible as well. For example, a gambling demon might measure and store the position of the molecule, but subsequently let a fair coin toss determine on which side he will attach the weight. If the weight goes up he subsequently follows the procedure of the regular demon, if the weight goes down he follows the procedure of the anti-demon. On average, half of the time work is obtained when the weight is lifted and half of the time work is wasted by letting the weight drop. The demon still requires work to operate his memory, so on average destroys work, albeit not as effective as the anti-demon. One might wonder why the gambling demon would measure the position of the molecule at all if that information does not play any role in where to put the weight. However, one might equally well ask: why not? Again, work and information are related only in an arbitrary way.

Of course, the anti-demon and gambling demon are contrived beings, rather different from what a rational physicist would do. Physicists typically do not like to waste work. Why assume that a demon does not want that too?

**4. Demons have goals too, but whence?**

It is clear from the above examples that the result of the demon's actions, and the possibility to connect information and work, depends on the implicit goal he has. If the demon is an automaton, a little machine, this goal has been imposed on it by how it was designed by the physicist. In effect, it follows her goal, not its own goal. But Maxwell's demon was originally conceived of by his creators as intelligent and with a will of his own. Maxwell speaks of "the intelligence of a very observant and neat-fingered being" (Maxwell in a 1867 letter to Tait, cited in Leff & Rex, 2003), and Thomson (1874) of "an intelligent being endowed with free will." Obviously, this is meant somewhat tongue-in-cheek, but it points to a real dichotomy. Either we assume that the demon is a mechanical device, that blindly follows the physicist's orders instilled in its design, or we assume that the demon is indeed acting on his own behalf. But in the latter case, why would he try to produce work rather than waste it? In other words, how could he acquire any consistent goal at all?[4]

A solution is suggested when we realize that Maxwell's demon appears to have properties normally ascribed only to living creatures, and if he is indeed alive he must be subject to evolution by natural selection. As I will argue below, an extended form of natural selection is sufficient, and presumably necessary, to provide demons with goals of their own. For constructing this case, several assumptions are made. The demons are assumed to be able to utilize free energy (potential work) from a rate-limited external source in order to operate the Szilard engine. Moreover, they can reproduce only when they operate the engine successfully, by first lifting the weight, and then reproduce through an unspecified mechanism driven by dropping the weight. Finally, their offspring randomly shifts (mutates) a little along the demon to anti-demon scale. Then the likes of those demons who waste as little work as possible will become more numerous, eventually.

A toy model (following van Hateren, 2015a) can illustrate this. Suppose that $n(h,t)$ is the number of demons of type $h$ at time $t$, and $\tau$ is the typical lifetime of a demon. Ignore mutation for the moment, and define a fitness $f(E,n,h)$ as the reproductive rate of the demon, properly normalized by (inverse) lifetime. The fitness is a function of $h$, but also of environmental circumstances $E(t)$ (e.g., shape and availability of cylinders, cords, pulleys, and weights) and of $n$ via the total number of demons $N=\Sigma_h n$ (with $\lim_{N\to\infty} f = 0$, because the external source of free energy will become exhausted when there are too many demons around). Then

$$\dot{n} + n(1-f)/\tau = 0 \qquad (1)$$

gives the dynamics of the population of demons. Demons of a type $h$ with $f>1$ will show exponential growth in numbers, and those with $f<1$ exponential decline. There is no asymptotic solution (apart from all demons becoming extinct), because it is assumed that $E$ keeps varying unpredictably, with a power-law spectral density (Bell, 2010; van Hateren, 2015a) such as $\sim 1/\omega^2$, with $\omega$ frequency.

Mutation is modelled as a convolution along the $h$-dimension with a weighting function such as a Gaussian $\Lambda$ with standard deviation $\lambda$. The convolution spreads the types of the offspring of a demon of type $h$ to neighbouring types, similar to diffusion along the $h$-dimension, with the speed of diffusion proportional to $\lambda$ (equivalent to a mutation rate[5]). We will assume here that $h$ determines how efficient the demon is in utilizing work, given the circumstances $E$. Thus there is no fixed $h$ that is always best in any circumstance, and a demon's line of descent[6] can therefore only survive if it continually adapts, by mutated offspring, to changing $E$. The dynamics is now given by (van Hateren 2015a)

$$\dot{n} + [n - \Lambda * (nf)]/\tau = 0, \qquad (2)$$

with $*$ denoting convolution along the $h$-dimension. Whereas in Eq. (1) the difference between $n$ and $nf$ determines whether $n$ increases or decreases for a particular $h$, in Eq. (2) it is the difference between $n$ and $\Lambda * (nf)$, because the latter is the new population distribution over $h$ produced by the mutations. Equation (2) describes the basic mechanism of evolution by natural selection as proposed by Darwin (1859). Mutation can be regarded, qualitatively, as a diffusion process that expands and explores $h$-space for suitable variants. Suitable is defined here as those types of demons that, given circumstances $E$, operate the Szilard engine successfully so that they can reproduce fast enough for letting their line of descent survive in the population.

Mutation in Eq. (2) operates with a fixed $\lambda$, implying a fixed mutation rate of demon offspring. However, a more optimal dynamics can be obtained (van Hateren, 2015a) by making $\lambda$ a function of an estimate of $f$, $\hat{f}(E,n,h)$, as made implicitly by the demons themselves and assumed to be a good estimator of the true fitness $f$. The dynamics is then given by

$$\dot{n}+[n-\Lambda(\hat{f})*(nf)]/\tau=0. \qquad (3)$$

---

[4] Obviously, if the demon were an automaton designed by a physicist, we could equally well ask how she could have acquired any consistent goal at all.
[5] Strictly speaking, the standard deviation of $\Lambda$ is $\lambda\Delta t$ in units of $h$, with $\Delta t=1$ the time step used for simulating Eqs. (2) and (3). An alternative would be to model the dynamics by modifying a regular diffusion equation.
[6] A sequence of demons related by direct descent; a demon's line of descent is singular in the past, but can split into multiple lines in the future, ranging from zero (extinction) to many.

The form of $\Lambda(\hat{f})$ needs to be special. The mutation rate $\lambda$ is decreased when the estimated fitness $\hat{f}$ is large, and increased when $\hat{f}$ is small. The result is that the offspring of demons with large fitness will not mutate much (small $\lambda$, i.e. slow diffusion along the $h$-dimension). This is a good strategy, because $E(t)$ typically changes slowly because of its power-law spectral density. Offspring with little change are then likely to have large fitness too, and thus keep the line of descent going. On the other hand, demons with small fitness are on the road to extinction. A way to avoid extinction is to have offspring which vary more (large $\lambda$, i.e. fast diffusion along the $h$-dimension). Most of this offspring will have even lower fitness than their parent, and they or their descendants are likely to be the end of their line of descent. But fast diffusion also increases the probability of finding or producing parts of $h$-space that yield fitness higher than average (again given $E$ at a particular time). Although this probability is small, it is potentially compensated by the exponential growth of reproduction when $f$ is sufficiently large. Simulations show that populations with $\lambda$ varying in this way outcompete populations with fixed $\lambda$ (van Hateren, 2015a). Interestingly, a range of molecular mechanisms that appear to conform to this scheme have recently been shown to exist in presumably all biological cells (reviewed in Galhardo et al., 2007).

The above scheme only works when $\hat{f}$ is a good estimate of $f$ and when $\Lambda(\hat{f})$ has a suitable form, with an inverse relationship between $\lambda$ and $\hat{f}$. But the forms of both $\Lambda(\hat{f})$ and $\hat{f}$ are also taken to be mutable and thus subjected to evolution by natural selection. Demons with forms of $\hat{f}$ producing a better estimate of the true fitness will outperform demons with less adequate $\hat{f}$. The dynamics of Eq. (3) is such that demons are, in effect, driven into the direction of both high $f$ and high $\hat{f}$. However, the two mechanisms have rather different signatures. High $f$ is obtained by a random search through $h$-space, whereas obtaining high $\hat{f}$ depends on modulating the search itself (by varying its variance $\lambda^2$) and also on how well $\hat{f}$ estimates $f$. An outcome of high $f$, as a consequence of the basic natural selection of Eq. (2), might appear to be the result of an implicit goal of the evolutionary process and the demons. But that would be merely an "as if" goal, only perceived after the fact. It is not different from other apparent, but not true goals one might perceive in other physical processes. For example, one might say that excited systems "tend to" fall back to lower energy states, or that low entropy systems "tend to" get higher entropy by dispersing to more microstates. When organisms "tend to" get or keep high fitness because of natural selection, this falls in the same category of these "as if" goals. In reality such processes just unfold as they do, often as a mere consequence of their statistics.

However, the interpretation of $\hat{f}$ as in Eq. (3) is different. Because $\lambda$ (i.e., $\Lambda$) is a function of $\hat{f}$ in a way that is known[7] to improve the expectation of the fitness of the demon's line of descent, obtaining high $f$ (and thus high $\hat{f}$) is a likely outcome of the dynamics. Importantly, $\hat{f}$ is an internal variable of the demon, a variable that is evolving and not fully determined, unlike $f$. Obtaining high $\hat{f}$ must therefore be regarded as a genuine goal of the demon and its line of descent. The demon's line of descent moves towards that goal along a trajectory that is the result of a feedback loop that entangles deterministic and stochastic factors (van Hateren, 2015a). The estimated fitness $\hat{f}$ drives the variance of offspring mutation, and the realized outcome of mutation determines $f$ and thus $\hat{f}$ of the next generation, subsequently driving the variance of mutation of their offspring, and so on and so forth.

---

[7] Implicitly known as instilled in genetic and molecular mechanisms established by previous natural selection, thus there is no true foresight involved, merely a probabilistic expectation.

Resulting trajectories are unpredictable in detail, but, on average, driven into the direction of the demon's goal, high $\hat{f}$. Equation (3) obviously does not provide the demon with intelligence, nor the consciousness required for free will (but see van Hateren, 2014, 2015a, for possibilities to construct more sophisticated demons). Nevertheless, it does produce a dynamics displaying at least the first primordial seeds of free will. Each new descendant seems to vary randomly, but is in fact partly determined by the entire history of both $\hat{f}$ and the realized mutations of the past line of descent. The spontaneity of randomness is thus combined with the goal-directedness of $\hat{f}$, producing a process that is better viewed as active rather than passive.

In conclusion, natural selection, as extended according to Eq. (3), can provide demons with true goals. Demons that reproduce as assumed here will indeed tend to operate the Szilard engine similarly to the demon of Fig. 1, consistently producing work (as a sub-goal of their true goal of reproducing). Moreover, they will drive those demons to extinction that lack true goals or that have goal-functions $\hat{f}$ inconsistent with $f$.

## 5. Functional information and formal information

In the previous section I argued that obtaining a high estimated fitness $\hat{f}$, as introduced in Eq. (3), is a true goal of a demon operating Szilard's engine in the way constructed here. High estimated fitness implies high true fitness, and high true fitness implies reproducing by operating Szilard's engine. This implies setting the memory required for operating the engine. In other words, setting the one-bit memory is a necessary sub-goal of the demon, because it is required for obtaining high fitness. The information in the one-bit memory is therefore not neutral, but goal-directed and functional for the demon. The information serves the demon's goal, namely to enable him to determine on which side of the cylinder the weight should be put, and thereby to lift the weight and reproduce.

Information is functional when it is expected to serve a genuine goal for the user of the information. Because the mechanism of Eq. (3) is presumably present in all forms of life (van Hateren, 2013), functional information is intrinsic to all forms of life. Functional information serves obtaining high $\hat{f}$, the ultimate goal of living organisms. For example, the DNA of a microbe can be regarded as representing functional information for the microbe, because it is expected to serve its $\hat{f}$ and thereby also $f$ (on average). Similarly, information operating within humans and within machines utilized by humans is also functional, because it is expected to promote its user's $\hat{f}$ and $f$, on average[8]. Information as used within living organisms or within technology used by humans might be called functional information. It is functional by definition whenever it is relevant for the organism's goals. Information of this kind is clearly a physical entity that really exists, because it has effects on the physical world independent of human theorizing and interpretation.

Landauer's general statement that information is physical depends on the notion that a physical quantity, work, is needed to store a bit of information in a memory. Conversely, a stored bit of information can be used in Szilard's engine to produce work. However, both conversions crucially depend on implicit goals. When storing a bit, the goal is given by the specified and therefore desired new state of the memory. If the new state were not specified, filling a memory with bits would cost no work at all. A stochastic, thermal process arbitrarily flipping the bits would do, at no cost. Conversely, a stored bit of information can only produce work in Szilard-type demons that have a specific goal, as argued above. Without such specific systems and goals, information cannot be used to produce work. In other words, information in general is not physical, only functional information is.

In addition to functional information, there is a second form of information, to be called formal information here, that does not imply any intrinsic goal-directedness but only depends on correlations. It is merely used as a tool of analysis, and only exists as theoretical, not as a physical entity with causal effectiveness. For example, methods derived from information theory can be used to analyse parts of the physical world that have no direct consequences for fitness (e.g. Enßlin et al., 2009). Such formal information is not functional in the sense defined above. It is just a conceptual tool that is useful when a particular system can be represented in such a way that it yields to the mathematical apparatus of information theory. But the information plays no causal role, independent of human analysis, within the system itself.

---

[8] For humans, fitness obviously is highly complex, incorporating cultural factors as well.

## 6. Are information and the second law of thermodynamics objective and universal?

Distinguishing functional and formal information opens a Pandora's box. As discussed above, information is strongly tied to entropy, as has long been argued by Jaynes (1957a, b). Moreover, Shannon entropy is basically the expectation value of a quantity that can be assigned to single states or events $i$, $\ln(1/p_i)$. This is a measure of how surprising an event is. It is large when a particular event has a small probability $p_i$ and zero when the event is certain. Information as obtained from Shannon entropy is, then, intimately tied to the probability of states or events, via a logarithmic transformation. Finally, heat and work as figuring in thermodynamics are also closely related to information, through Landauer's principle as discussed above. The implicit goal-directedness of functional information corresponds to the explicit technological goal-directedness of heat, work, and the second law. A primary aim of thermodynamics in the era of the steam engine was to understand the relationship between useful energy (work) and useless energy (heat). The second law basically says that potentially useful energy has a tendency to disperse and become useless, i.e. useless as a source of work serving living organisms (see also Myrvold, 2011). The above connections imply that when information comes in two kinds, functional (physically effective) and formal (as a theoretical tool), this must be true also for entropy, probability, and the second law.

    Does this mean that information and the related concepts (abbreviated to "information" below) are subjective quantities, depending on the contingencies of individual human thought[9]? Fortunately not. The first thing to note is that information even in its functional variant is not limited to humans, but is intrinsic to all forms of life and their devices. Moreover, the goal-directedness of functional information is an objective and universal property of the systems involved. For example, the true fitness and the estimated fitness (to the extent that it is indeed close to the true fitness) are objective properties of a microbe from the point of view of the microbe and its line of descent. The difference between survival and extinction is objective, not subjective, hence also the information that determines this difference is objective. Functional information is even universal, because there is only one way to understand fitness from an external point of view, i.e. the point of view of an intelligent being studying microbes. Also information in technology can be analysed objectively and universally, despite the fact that technology does not have intrinsic goals, but only human-defined ones. But information is objective and universal for the latter, and therefore also for technology serving human goals.

    However, the case of information in its formal, causally ineffective variant is different. When defining information for a system that does not have intrinsic goals, there is some arbitrariness in how the system is parsed and subdivided in order to attribute information to the system. Thus there is no universality. But that does not imply that it is subjective. Parsing and subdividing can be subjected to further criteria, such as simplicity and solvability, that may leave only one or a few options open. Moreover, once a subdivision is made, theoretical derivations and empirical measurements can be agreed upon as objectively valid by any intelligent being. It is the prerogative of the investigator to conceive of the structure of a system in such a way that it is analysable, i.e. such that an existing body of theoretical and experimental techniques can be applied to it. Nevertheless, the lack of universality when ascribing information to entities outside the realm of life and technology is reason for caution.

---

[9] Or, in the case of intersubjectivity, depending on the contingencies of collective human thought.

Caution is needed when assuming that associated general laws, such as the second law, can be extrapolated beyond the realm where they have originated and have been well established[10].

## 7. Conclusion

As noted by Leff and Rex (2003), Maxwell's demon seems to be cat-like, not in the sense that he does not know if he is dead or alive, but in the sense that he returns alive and kicking each time when he has been presumed dead. Perhaps this resilience is explained by his life-like ability to let mutated progeny return in his stead, apparently to haunt generations of physicists. However, the presumed goal of the demons considered in this article is not to haunt physicists, but rather to procreate by using Szilard's one-molecule engine. By doing that, the information utilized for operating the engine becomes functional in the sense of being necessary for the demon's goal, and thereby acquires physical reality. The consequence is that information, as well as entropy, probability, and the second law, comes in two kinds. The first, functional kind is physical and causally effective, and it is the one associated with life and technology. The second, formal kind is only theoretical, and could also be attributed to anything lacking intrinsic or derived goals. Whereas the first kind is objective, universal, and functional, the second kind is only objective.

That at least some forms of entropy and probability strongly depend on human analysis has been argued before, most notably by Jaynes (1957a, b; 2003). Moreover, uncertainty about the interpretation and the extent of the second law has a long history (Uffink, 2001; Čápek & Sheenan, 2005). The theory presented here suggests an important role for the emergence of life, as life appears to be the only source of true goals. A tight connection between biology and information has been noticed before. For example, Schneider (2006) argues that Shannon's information theory is essentially a theory about biological systems, because it presupposes the ability to categorize and make choices, which is a biological property. Kauffman (2003) questions the criteria for defining work when he discusses life, because objective criteria for defining structure and organization are not clear, which is implicitly also a problem of choosing. The current theory suggests that the goal-directedness implied by Eq. (3) and its extensions (van Hateren, 2013, 2014, 2015a,b) provides an explanation for these considerations and similar ones. It also suggests that recent proposals that information as such is physical and has a constitutive role for the emergence of life (Walker & Davies, 2012), of consciousness (Tononi, 2008), or even of the entire universe (Wheeler, 1989) have got it backwards. Information is a derived property of life, human thinking, and scientific analysis, not the other way around.

**References**

Belfer, I. (2012). The info-computation turn in physics. In A. Voronkov (Ed.), *Turing-100. The Alan Turing Centenary. EPiC Series 10* (pp. 20-33). EasyChair.

Belfer, I. (2014). Informing physics: Jacob Bekenstein and the informational turn in theoretical physics. *Physics in Perspective,* 16, 69–97.

Bell, G. (2010). Fluctuating selection: the perpetual renewal of adaptation in variable environments. *Philosophical Transactions of the Royal Society B, 365*, 87-97.

Bennett, C. H. (1982). The thermodynamics of computation – a review. *International Journal of Theoretical Physics, 21*, 905–940.

Čápek, V., & Sheenan D. P. (2005) *Challenges to the second law of thermodynamics: theory and experiment.* Dordrecht: Springer.

Darwin, C. (1859) *On the origin of species by means of natural selection.* London: John Murray.

---

[10] Specifically, the second law has been well established for the typical circumstances of life, i.e., molecular systems in near-constant gravity and with negligible long-range electromagnetic and gravitational interactions.

Enßlin, T. A., Frommert, M., & Kitaura, F. S. (2009). Information field theory for cosmological perturbation reconstruction and nonlinear signal analysis. *Physical Review D, 80*, 105005

Galhardo, R. S., Hastings, P. J., & Rosenberg, S. M. (2007). Mutation as a stress response and the regulation of evolvability. *Critical Reviews in Biochemistry and Molecular Biology, 42*, 399-435.

Jarzynski, C. (2011). Equalities and inequalities: irreversibility and the second law of thermodynamics at the nanoscale. *Annual Review of Condensed Matter Physics, 2*, 329-351.

Jaynes, E. T. (1957a). Information theory and statistical mechanics. *Physical Review, 106*, 620-630.

Jaynes, E. T. (1957b). Information theory and statistical mechanics. II. *Physical Review, 108*, 171-190.

Jaynes, E. T. (2003). *Probability theory: the logic of science.* Cambridge: Cambridge University Press.

Kauffman, S. A. (2003). Molecular autonomous agents. *Philosophical Transactions of the Royal Society A, 361*, 1089-1099.

Landauer, R. (1961). Irreversibility and heat generation in the computing process. *IBM Journal of Research and Development, 5*, 183–191.

Landauer, R. (1991). Information is physical. *Physics Today, 44(5)*, 23-29.

Leff, H. S. (1999). What if entropy were dimensionless? *American Journal of Physics, 67*, 1114-1122.

Leff, H. S., & Rex, A. F. (2003). *Maxwell's demon 2: entropy, classical and quantum information, computing.* Bristol: IOP publishing.

Maruyama, K., Nori, F., & Vedral, V. (2009). Colloquium: the physics of Maxwell's demon and information. *Reviews of Modern Physics, 81*, 1-23.

Maxwell, J. C. (1871). *Theory of heat.* London: Longmans, Green, and Co.

Myrvold, W. C. (2011). Statistical mechanics and thermodynamics: a Maxwellian view. *Studies in the History and Philosophy of Modern Physics, 42*, 237-243.

Robinson, L., Bawden, D. (2014). Mind the gap: transitions between concepts of information in varied domains. In F. Ibekwe-SanJuan, & T. M. Dousa (Eds.), *Theories of information, communication and knowledge* (pp. 121-141). Dordrecht: Springer.

Schneider, T. D. (2006). Claude Shannon: biologist. *IEEE Engineering in Medicine and Biology Magazine, 25*, 30-33.

Shannon, C. E. (1948). A mathematical theory of communication. *Bell System Technical Journal, 27*, 379–423.

Szilard, L. (1929). Über die Entropieverminderung in einem thermodynamischen System bei Eingriffen intelligenter Wesen [On the decrease of entropy in a thermodynamic system by the intervention of intelligent beings]. *Zeitschrift für Physik, 53*, 840–856.

Thomson, W. (1874). Kinetic theory of the dissipation of energy. *Nature, 9*, 441-444; reprinted in Leff and Rex, 2003, Chapter 2.1.

Tononi, G. (2008). Consciousness as integrated information: a provisional manifesto. *Biological Bulletin, 215*, 216-242.

Uffink, J. (2001). Bluff your way in the second law of thermodynamics. *Studies in the History and Philosophy of Modern Physics, 32*, 305-394.

van Hateren, J. H. (2013). A new criterion for demarcating life from non-life. *Origins of Life and Evolution of Biospheres, 43*, 491-500. doi:10.1007/s11084-013-9352-3

van Hateren, J. H. (2014). The origin of agency, consciousness, and free will. Phenomenology and the Cognitive Sciences. doi:10.1007/s11097-014-9396-5; https://sites.google.com/site/jhvanhateren/home/agency/agency_web.pdf

van Hateren, J. H. (2015a). Active causation and the origin of meaning. *Biological Cybernetics, 109*, 33-46. doi:10.1007/s00422-014-0622-6;  arXiv:1310.2063

van Hateren, J. H. (2015b). Biological function and functional information. In preparation

Walker, S. I., & Davies, P. C. W. (2012). The algorithmic origins of life. *Journal of the Royal Society Interface, 10*, 20120869.

Wheeler, J. A. (1989). Information, physics, quantum: the search for links. *Proceedings 3rd International Symposium on Foundations of Quantum Mechanics,* Tokyo (pp. 354-368).